\documentclass[12pt]{article}

\usepackage{amssymb}
\usepackage{amsmath, amsthm}
\usepackage{amscd}
\usepackage{latexsym}
\usepackage{graphicx}
\usepackage[all]{xy}
\input xy
\xyoption{all}
\usepackage{eucal}

 \makeatletter
\newfont{\footsc}{cmcsc10 at 8truept}
\newfont{\footbf}{cmbx10 at 8truept}
\newfont{\footrm}{cmr10 at 10truept}
\renewcommand{\ps@plain}{%
\renewcommand{\@oddfoot}{\footsc Spitzer's identity and the algebraic Birkhoff
decomposition in pQFT, {\footbf July 11,
2004},\hfil\footrm\thepage}} \makeatother 

\begin{document}

\def\mkb{\mbox}
\def\beq{\begin{equation}}
\def\eeq{\end{equation}}
\def\beqn{\begin{eqnarray}}
\def\eeqn{\end{eqnarray}}

\newcommand{\mfootnote}[1]{{}}                          
\newcommand{\mlabel}[1]{\label{#1}}                     
\newcommand{\mcite}[1]{\cite{#1}}                       
\newcommand{\mref}[1]{Eq. (\ref{#1})}

\newtheorem{theorem}{Theorem}[section]
\newtheorem{proposition}[theorem]{Proposition}
\newtheorem{definition}[theorem]{Definition}
\newtheorem{lemma}[theorem]{Lemma}
\newtheorem{corollary}[theorem]{Corollary}
\newtheorem{prop-def}{Proposition-Definition}[section]
\newtheorem{claim}{Claim}[section]
\newtheorem{remark}[theorem]{Remark}
\newtheorem{example}[theorem]{Example}
\newtheorem{propprop}{Proposed Proposition}[section]
\newtheorem{conjecture}{Conjecture}
\newenvironment{exam}{\begin{example}\rm}{\end{example}}
\newenvironment{rmk}{\begin{remark}\rm}{\end{remark}}


%
\def\ta1{\includegraphics[scale=0.42]{tree1}}
\def\tb2{\includegraphics[scale=0.42]{tree2}}
\def\tc3{\includegraphics[scale=0.42]{tree3}}
\def\td31{\!\!\includegraphics[scale=0.42]{tree31}}
\def\te4{\includegraphics[scale=0.42]{tree4}}
\def\tf41{\!\!\includegraphics[scale=0.42]{tree41}}
\def\tg42{\!\!\includegraphics[scale=0.42]{tree42}}
\def\th43{\!\!\includegraphics[scale=0.42]{tree43}}
\def\ti5{\includegraphics[scale=0.42]{tree5}}
\def\tj51{\!\!\includegraphics[scale=0.42]{tree51}}
\def\tk52{\!\!\includegraphics[scale=0.42]{tree52}}
\def\tl53{\!\!\includegraphics[scale=0.42]{tree53}}
\def\tm54{\!\!\includegraphics[scale=0.42]{tree54}}
\def\tn55{\!\!\includegraphics[scale=0.42]{tree55}}
\def\tp56{\!\!\includegraphics[scale=0.42]{tree56}}
\def\tq57{\!\!\includegraphics[scale=0.42]{tree57}}
\def\tr58{\!\!\includegraphics[scale=0.42]{tree58}}
%
\def\dta{\includegraphics[scale=0.42]{dectree1}}
\def\dtb{\includegraphics[scale=0.42]{dectree2}}
\def\dtba{\includegraphics[scale=0.42]{dectree21}}
\def\dtab{\includegraphics[scale=0.42]{dectree12}}
\def\graph1{\includegraphics[scale=0.42]{graph1}}
%
%
\def\sdta{\includegraphics[scale=0.62]{dectree1}}
\def\sdtb{\includegraphics[scale=0.62]{dectree2}}
\def\sta1{\includegraphics[scale=0.62]{tree1}}
\def\stb2{\includegraphics[scale=0.62]{tree2}}
\def\stc3{\includegraphics[scale=0.62]{tree3}}
\def\std31{\!\!\includegraphics[scale=0.62]{tree31}}
\def\ste4{\includegraphics[scale=0.62]{tree4}}
\def\stf41{\!\!\includegraphics[scale=0.62]{tree41}}
\def\stg42{\!\!\includegraphics[scale=0.62]{tree42}}
\def\sth43{\!\!\includegraphics[scale=0.62]{tree43}}
\def\sti5{\includegraphics[scale=0.62]{tree5}}
\def\stj51{\!\!\includegraphics[scale=0.62]{tree51}}
\def\stk52{\!\!\includegraphics[scale=0.62]{tree52}}
\def\stl53{\!\!\includegraphics[scale=0.62]{tree53}}
\def\stm54{\!\!\includegraphics[scale=0.62]{tree54}}
\def\stn55{\!\!\includegraphics[scale=0.62]{tree55}}
\def\stp56{\!\!\includegraphics[scale=0.62]{tree56}}
\def\stq57{\!\!\includegraphics[scale=0.62]{tree57}}
\def\str58{\!\!\includegraphics[scale=0.62]{tree58}}
%
%
\def\A{\mathcal{A}}
\def\B{{B}^{+}}
\def\BR{\mathcal{B}}
\def\D{\mathcal{D}}
\def\F{\mathcal{F}}
\def\G{\mathcal{G}}
\def\H{\mathcal{H}}
\def\L{\mathcal{L}}
\def\M{\mathcal{M}}
\def\P{\mathcal{P}}
\def\RB{\mathcal{R}}
\def\T{\mathcal{T}}
\def\U{\mathcal{U}}
\def\X{\mathcal{X}}
\def\l{\mathfrak{l}}
\def\g{\mathfrak{g}}
\def\C{\mathbb{C}}
\def\K{\mathbb{K}}
\def\N{\mathbb{N}}
\def\Q{\mathbb{Q}}
\def\R{\mathbb{R}}
\def\Z{\mathbb{Z}}
\def\to{\rightarrow}
\def\lto{\longrightarrow}
\def\sh{\sqcup \!\! \sqcup}
\def\EXP{exp^{\star}}
\def\EXPR{exp^{\star_R}}
\def\ep{\epsilon}
\def\CH{char_{\A}\H_{rt}}
\def\CHR{char_{\A_R}\H_{rt}}
\def\pCH{\partial char_{\A}\H_{rt}}
\def\pCHR{\partial \char_{\A_R}\H_{rt}}
\def\Lie{\mathcal{L}_{\H_{rt}}}
\def\id{\mathrm{id}}
\def\mchar{\mathrm{char}}


\begin{center}
{\LARGE{Spitzer's Identity and the Algebraic Birkhoff Decomposition in pQFT}}\\[1.5cm]
\end{center}

\begin{center}
         KURUSCH EBRAHIMI-FARD\footnote{kurusch@ihes.fr}\\\smallskip
{\small{
         Institut Henri Poincar\'{e}\\
         11, rue Pierre et Marie Curie\\
         F-75231 Paris Cedex 05, France\\
         {\small{and}}\\
         Universit\"at Bonn -
         Physikalisches Institut\\
         Nussallee 12,
         D-53115 Bonn, Germany}}\\[0.6cm]

         LI GUO\footnote{liguo@newark.rutgers.edu}\\\smallskip
{\small{
         Rutgers University\\
         Department of Mathematics and Computer Science\\
         Newark, NJ 07102, USA}}\\[0.6cm]

         DIRK KREIMER\footnote{kreimer@ihes.fr and dkreimer@bu.edu, Center for Math.Phys.,
         Boston University.}\\\smallskip
{\small{
         C.N.R.S.-I.H.\'E.S.\\
         Le Bois-Marie, 35, Route de Chartres\\
         F-91440 Bures-sur-Yvette, France}}\\[0.3cm]
\end{center}

\begin{center}
July 11, 2004\\[1.5cm]
\end{center}

\begin{abstract}
In this article we continue to explore the notion of Rota-Baxter
algebras in the context of the Hopf algebraic approach to
renormalization theory in perturbative quantum field theory. We
show in very simple algebraic terms that the solutions of the
recursively defined formulae for the Birkhoff factorization of
regularized Hopf algebra characters, i.e.\ Feynman rules,
naturally give a non-commutative generalization of the well-known
Spitzer's identity. The underlying abstract algebraic structure is
analyzed in terms of complete filtered Rota-Baxter algebras.\\[0.5cm]
\end{abstract}


\section{Introduction}

The theory of Rota-Baxter type algebras has a long and interesting
history. It was introduced by the American mathematician Glen
Baxter in 1960 \cite{Baxter} in the context of fluctuations in
probability theory. The subject was further explored especially by
F. V. Atkinson \cite{A}, J. F. C. Kingman \cite{Kin}, P. Cartier
\cite{C} and others, but foremost by the mathematician Gian-Carlo
Rota in his work in the late 1960s and early 1970s \cite{RSmith,
Rota1} and later in his beautiful reviews \cite{Rota2, Rota3}. In
the center of these works stood the category of commutative
associative Rota-Baxter algebras and its free objects. Recently,
one of us together with W. Keigher gave a very concise description
of the latter in terms of a mixable shuffle product \cite{G-K1,
G-K2}, which provides a generalization of the classical shuffle
product \cite{E-G2}. As one of the main results of the above early
work on (free) commutative Rota-Baxter algebras simple
combinatorial and analytical proofs of Spitzer's identity were
obtained. The latter is in its classical form a well-known object
in probability theory having many applications.

Let us mention here that the Lie algebraic version of the
Rota-Baxter relation plays under the name (modified) classical
Yang-Baxter\footnote{Here the relation is named after the
physicists C.-N.Yang and Rodney Baxter.} equation a prominent
r\^ole in the theory of integrable systems \cite{BBT, STS2, STS1,
STS3}. Classical R-matrices, i.e. solutions of this equation, are
connected to the Riemann-Hilbert problem and related factorization
problems.

Recently the notion of Rota-Baxter algebra reappeared in the
mathematics and, above all, physics literature. On the mathematics
side we would like to underline its intimate link to Loday's
dendriform algebra structures \cite{Agu1, E-G1, KEF, PhL1, JLL1}.
From a physics viewpoint it appeared in the Hopf algebraic
approach to the theory of renormalization in perturbative quantum
field theory (pQFT). This approach provided a solid mathematical
frame for renormalization theory in terms of combinatorial Hopf
algebras of Feynman graphs \cite{CK1, CK2, Kreim1, Kreim2}.

Here, we will dwell mainly on the latter aspect by showing that
the recently given solutions to the recursively defined formulae
for the algebraic Birkhoff factorization of regularized Hopf
algebra characters in terms of a so-called BCH-recursion
\cite{EGK1, EGK2} provide natural non-commutative generalizations
of the above mentioned Spitzer's identity. We introduce the notion
of complete filtered not necessarily commutative Rota-Baxter
algebras to underline the abstract algebraic structure giving rise
to this factorization. This approach allows us to derive in a
fairly simple manner an alternative recursion for renormalized
Feynman rules.

The paper is organized as follows. In Section \ref{Rota-Baxter} we
collect some basic facts about (not necessarily commutative)
Rota-Baxter algebras. Section \ref{Sect-spitzer} contains the
non-commutative generalization of Spitzer's identity in the
context of complete filtered Rota-Baxter algebras and an abstract
algebraic formulation of Bogoliubov's recursion. Using the above
results we finish this paper with a short review of the Birkhoff
decomposition of regularized Hopf algebra characters. This turns
out to be an example for the more general content of the forgoing
section, placed in the context of the Hopf algebra approach to
renormalization theory. It allows us to derive a new recursion
formula for renormalized Feynman rules solely based on iterating
the renormalized character $\phi_+$ instead of the countertem
$\phi_-$. We finish this article with a short summary and outlook.

\section{Rota-Baxter algebras}\label{Rota-Baxter}

Let $\K$ be a field of characteristic $0$. By a $\K$-algebra we
mean an associative algebra over $\K$ that is not necessarily
unital nor commutative unless stated otherwise.

\begin{definition}
Let $\A$ be a $\K$-algebra with a $\K$-linear map $R: \A \to \A$.
We call $\A$ a Rota-Baxter $\K$-algebra and $R$ a Rota-Baxter map
(of weight $\theta  \in \K$) if the operator $R$ holds the
following Rota-Baxter relation of weight $\theta \in \K$
\footnote{Some authors denote this relation in the form
$R(x)R(y)=R\big(R(x)y + xR(y)+\lambda xy\big)$. So that $\lambda =
-\theta$.}:
\begin{equation}
    R(x)R(y) + \theta R(xy) = R\big(R(x)y + xR(y)),\ \forall x,y \in \A.
    \label{RBR}
\end{equation}
\end{definition}

\begin{rmk}
{\rm{ (0) Obviously, the above definition extends to
non-associative algebras in general, and the field $\K$ may be
replaced by an arbitrary commutative ring.
\smallskip

(1) In the rest of the paper we will fix the weight $\theta=1$,
which is called the standard form or the Rota-Baxter relation.
\smallskip

(2) If $R$ fulfills the standard form of (\ref{RBR}) then
$\tilde{R}:=id_{\A}-R$ fulfills the same Rota-Baxter relation.
\smallskip

(3) The ranges of $R$ and $\tilde{R}=id_{\A}-R$ give subalgebras
in $\A$.
\smallskip

(4) $R$ and $\tilde{R}=id_{\A}-R$ fulfill the following mixed
relations
\allowdisplaybreaks{
\begin{eqnarray}
R(x)\tilde{R}(y) &=& R\big(x \tilde{R}(y)\big) +
\tilde{R}\big(R(x)y\big) \label{ab}\\
\tilde{R}(x) R(y)&=& R\big(\tilde{R}(x)y\big) +
\tilde{R}\big(xR(y)\big), \;\; x,y \in \A. \label{ba}
\end{eqnarray}}}}
\end{rmk}

\begin{exam}
{\rm{ (0) On the algebra of Laurent series $\C[\epsilon^{-1},
\epsilon]]$ we have (up to automorphisms,) only the following two
Rota-Baxter maps $R^{(r)},\; r=0,1$. Both are of weight $\theta=1$
and defined as follows. For $\sum_{k=-m}^{\infty}c_k\epsilon^{k}
\in \C[\epsilon^{-1},\epsilon]]$ they give:
\begin{equation}
    R^{(r)} \big(\sum_{k = -m}^{\infty}c_k\epsilon^{k}\big):=\sum_{k=-m}^{-r}c_k\epsilon^{k},\; r=0,1.
    \label{Rms}
\end{equation}
Within renormalization theory, so-called dimensional
regularization together with the minimal subtraction scheme, i.e.
$R_{MS}:=R^{(1)}$, play an important r{\^o}le \cite{Kreim2}.
\smallskip

(1) The case of a Rota-Baxter map of weight $\theta=0$, i.e.
$R(x)R(y) = R\big(R(x)y + xR(y)\big)$, naturally translates into
the ordinary shuffle relation, and finds its most prominent
example in the integration by parts rule for the Riemann integral.
On the other hand Jackson's q-integral \cite{Rota2} gives a
generalization of the Riemann integral to a Rota-Baxter map of
weight $\theta=1-q$.}}
\end{exam}

\begin{proposition}
In the case of the Rota-Baxter algebra $\A$ to be a Lie admissible
$\K$-algebra, the Rota-Baxter relation naturally extends to the
Lie algebra $\L_{\A}$ with commutator bracket $[x,y]:=xy-yx,\;
\forall x,y \in \A$:
\begin{equation}
    \label{LieRBR}  [R(x),R(y)] +  R([x,y]) = R\big([R(x),y] +
    [x,R(y)]\big).
\end{equation}
\end{proposition}

\begin{proposition}\label{prop-double}
Let $\A$ be a Rota-Baxter algebra with Rota-Baxter map $R$.
Equipped with the new product

\begin{equation}
  \label{double}  a \ast_R b := R(a)b + aR(b) - ab,             \\
\end{equation}

the vector space underlying $\A$ is again a Rota-Baxter algebra of
the same type, denoted by $\A_R$.
\end{proposition}

The proof of this Proposition is a fairly easy exercise and
follows directly from the identity (\ref{RBR}) for $\theta=1$. We
call this new Rota-Baxter algebra $(\A_R,R)$ the double of $\A$,
and $\ast_R$ the double product.

\begin{rmk}
{\rm{ (0) Let us remark here that this double construction
appeared in a Lie algebraic context in \cite{STS1}, where the name
was coined.
\smallskip

(1) The product $\ast_R$ can be written using $R$ and
$\tilde{R}=id_{\A}-R$:
\begin{equation}
a \ast_R b = R(a)b - a\tilde{R}(b), \label{double2}
\end{equation}
which can be interpreted in terms of the dendriform dialgebra
structure of Loday \cite{JLL1}.

(2) From the definition of the $\ast_R$ product in (\ref{double})
it is obvious, that $R$ and $\tilde{R}=id_{\A}-R$ become an (not
necessarily unital) algebra homomorphism and anti-homomorphism,
respectively, from the double $\A_R$ to $\A$: \allowdisplaybreaks{
\begin{eqnarray}
     R(a \ast_R b) &=& R(a)R(b) \label{Rhom1} \\
  \tilde{R}(a \ast_R b)&=& -\tilde{R}(a) \tilde{R}(b). \label{Rhom2}
\end{eqnarray}}
(3) As $(\A_R,\ast_R)$ is again of Rota-Baxter type, the above
construction of the double extends to the so-called Rota-Baxter
double hierarchy \cite{EGK2}.}}
\end{rmk}

By definition, for the double product (\ref{double}) we have
$$
a \ast_R b = R(a)R(b)-\tilde{R}(a)\tilde{R}(b)
$$
and so by (\ref{Rhom1}),
$$
R(a)R(b)=R\big(R(a)R(b)-\tilde{R}(a)\tilde{R}(b)\big).
$$
Inductively, this can be generalized to
\allowdisplaybreaks{
\begin{equation}
\prod_{i=1}^{n} R(x_i) = R\bigg( \prod_{i=1}^{n} R(x_i) -
(-1)^{n}\prod_{i=1}^{n} \tilde{R}(x_i)\bigg), \;\; x_i \in \A,\;
i=1 \dots n \label{Kingman}
\end{equation}}
and then specialized to the following simple formula of Kingman
which appeared in~\cite{Kin}.
\begin{equation}
R(u)^n = R\big(R(u)^n- (-\tilde{R}(u))^n\big ),\;\; u\in \A.
\mlabel{eq:Kingman}
\end{equation}


\section{Non-commutative Spitzer's formula}\label{Sect-spitzer}

In the following, we do assume that an algebra in general is
associative and unital, the unit will be denoted by $1$, but we do
not assume that the algebra is commutative.

\subsection{Spitzer's formula}

Spitzer's formula~\cite{Sp} is regarded as a remarkable stepping
stone in the theory of sums of independent random variables in the
fluctuation theory of probability. It was also the motivation for
Baxter to define his identity \cite{Baxter}. The identity of
Spitzer has the following algebraic formulation.

\begin{theorem} {\rm \cite{RSmith}}
Let $(\A,R)$ be an unital commutative Rota-Baxter $\Q$-algebra of
weight $\theta = 1$. Then for $a \in \A$, we have
\begin{equation}
 \exp\left (R(\log(1-ax)^{-1}) \right )
    =\sum_{n=0}^\infty x^n \underbrace{R\big( R( R( \cdots (R(a)a ) a ) a) \big)}_{n\mbox{\rm -}{\rm times}}
    \mlabel{eq:si1}
\end{equation}
in the ring of power series $\A[[x]]$.
\end{theorem}
For other than the combinatorial proofs of Spitzer and Baxter, we
refer the interested reader to see \cite{A, C, Kin, RSmith, We}.

Using our previous work~\cite{EGK2} on the Birkhoff decomposition
of regularized characters in the Hopf algebraic approach to
renormalization theory in pQFT, we will derive a non-commutative
version of Spitzer's formula. Quite remarkably, the proof
presented here is similar to the one given in the commutative case
by Kingman~\cite{Kin}. Furthermore, once this formula is obtained,
a simple but beautiful result of Atkinson in
Theorem~\ref{Atkinson2} applies to give us a new recursive formula
back in the realm of Birkhoff decomposition in renormalization
theory with respect to the so-called renormalized character, which
we will describe in the next section.

We first consider Rota-Baxter algebras with a complete filtration.
This setup allows us to apply our results to the Rota-Baxter
algebra of renormalization introduced in \cite{EGK1}. The general
case of Rota-Baxter algebras $\A$ will be treated by considering
the power series ring $\A[[x]]$ in the commuting variable $x$.

\subsection{Complete Rota-Baxter algebras}
We first introduce the category of complete Rota-Baxter algebras.

\begin{definition}
A {\bf filtered Rota-Baxter algebra} is a Rota-Baxter algebra
$(\A,R)$ together with a decreasing filtration $\A_n, \: n\geq 0$
of Rota-Baxter subalgebras. Thus we have
$$
\A_n \A_m \subseteq \A_{n+m}
$$
and
$$
R(\A_n) \subseteq \A_n.
$$
Such a filtered Rota-Baxter algebra is called {\bf
complete}\footnote{To avoid possible confusions, we alert the
reader that in \cite{G-K2} the concept of complete filtered
Rota-Baxter algebras has been defined, where the filtration is
canonically derived from the Rota-Baxter operator. That definition
is not needed in this paper.}
 if
$\cap \A_n = 0$ and if the resulting embedding
$$
\A \to \bar{\A}:= \varprojlim \A/\A_n
$$
is an isomorphism.
\end{definition}

By the completeness of the filtered Rota-Baxter algebra $(\A,R)$,
the functions
$$
\exp: \A_1 \to 1+\A_1,\ \exp(a):=\sum_{n=0}^\infty \frac{a^n}{n!},
$$
$$
\log: 1+\A_1 \to \A_1,\ \log(1+a):=-\sum_{n=1}^\infty \frac{(-a)^n}{n}
$$
are well-defined.
This has the following (classical) interpretation of Lie groups and Lie
algebras.

$1+\A_1$ has a Lie group structure by the multiplication in $\A$,
and $\A_1$ has a Lie algebra structure by the commutator bracket
$[a,b]:=ab-ba$. Then the maps $\exp$ and $\log$ are the
isomorphisms from the Lie algebra to the Lie group and its
inverse.

\begin{exam} \label{FG}
For the Hopf algebra $\H_{FG}$ of Feynman graphs (or rooted trees)
and the ring of Laurent series $\A:=\C[\ep^{-1},\ep]]$ with the
Rota-Baxter operator defined to be the projection to the pole
part, i.e. $R:=R^{(1)}:\C[\ep^{-1},\ep]] \to
\epsilon^{-1}\C[\ep^{-1}]$ in (\ref{Rms}), the algebra
$L(\H_{FG},\A)$ with the convolution product and lifted
Rota-Baxter map $\RB : L(\H_{FG},\A) \to L(\H_{FG},\A)$ is a
complete Rota-Baxter algebra \cite{CK1, FGV, Kreim1}.
See~\cite[II.3.3.]{M} for the proof. Further in this setting $1 +
A_1$ is the group of (regularized) characters and $A_1$ is the Lie
algebra of infinitesimal characters.
\end{exam}

\smallskip
For $a \in \A$, inductively define
$$
(R a)^{[n+1]}:=R\big((Ra)^{[n]}\:a\big)\;\; \makebox{and} \;\;(R
a)^{\{n+1\}}:=R\big(a\:(Ra)^{\{n\}}\big)
$$
with the convention that $(Ra)^{[1]}=R(a)=(Ra)^{\{1\}}$ and
$(Ra)^{[0]}=1=(Ra)^{\{0\}}$.

Also by the completeness, there is a unique map $\chi: \A_1 \to
\A_1$ that satisfies the equation
\begin{equation}
\chi(a)=a - BCH\big(R(\chi(a)),\tilde{R}(\chi(a))\big)
\label{BCH-recur}
\end{equation}
which was introduced in \cite{EGK2} and will be coined as $BCH$-recursion for short.
Here $BCH(x,y)$ denotes the Baker-Campbell-Hausdorff formula such
that
$$
\exp(x)\exp(y)=\exp\big(x+y+BCH(x,y)\big)
$$
which is a power series in $x,y$ of degree 2. Relation
(\ref{BCH-recur}) was used in our approach to the algebraic
Birkhoff factorization, in connection with a classical R-matrix
notion coming from a Lie Rota-Baxter relation (\ref{LieRBR}), see
Section \ref{Birkhoff}.

We call it the $BCH$-recursion since $\chi(a)$ is defined to be
$\lim_{n \to \infty} \chi_n(a)$ where
\allowdisplaybreaks{\begin{eqnarray*}
\chi_0(a) &=& a, \\
\chi_{n+1}(a) &=& a - BCH\big( R(\chi_n(a)),\tilde{R}(\chi_n(a))
\big).
\end{eqnarray*}}
To see why this gives the unique solution to recursion relation
(\ref{BCH-recur}), we first define for $a \in \A$, $\Lambda : \A
\to \A$ \cite{EGK2}
$$
\Lambda(a):=BCH\big(R(a),\tilde{R}(a)\big).
$$
Then for $s \in A_n, n \geq 1$, $\Lambda(a+s)$ is $\Lambda(a)$ plus
a sum in which each term has $s$ occurring at least once, and
hence is contained in $A_{n+1}$. Thus we have
\begin{equation}
 \Lambda(a \mod A_n) \equiv \Lambda(a) \mod A_{n+1}.
\mlabel{eq:cong}
\end{equation}
Now we have
$$
\chi_1(a)= a + \Lambda(\chi_0(a)) = a + \Lambda(a) \equiv a \equiv
\chi_0(a) \mod A_{2}.
$$
By induction on $n$ and (\ref{eq:cong}), we have
\begin{eqnarray*}
 \chi_{n+1}(a)& =& a + \Lambda(\chi_n(a)) \\
&\equiv & a + \Lambda(\chi_{n-1}(a) \mod A_{n+1}) \\
& \equiv & a + \Lambda(\chi_{n-1}(a)) \mod A_{n+2}\\
& \equiv & \chi_n(a) \mod A_{n+2}.
\end{eqnarray*}
Thus $\lim_{n\to \infty} \chi_n(a)$ exists and is a solution of (\ref{BCH-recur}).

Suppose $b$ is another solution. Then, as above, we have
$$\chi_0(a) = a \equiv a + \Lambda(b) \equiv b \mod A_2.$$
Induction on $n$ gives the following
\begin{eqnarray*}
\chi_{n+1}(a) &=& a+\Lambda(\chi_n(a)) \\
    &\equiv & a + \Lambda(b \mod A_{n+2})\\
    &\equiv & a + \Lambda(b) \mod A_{n+3}\\
    &\equiv & b \mod A_{n+3}.
\end{eqnarray*}
Thus $b=\lim_{n\to \infty} \chi_n(a).$
The reader may find it helpful to consult the nice expository work of
Manchon \cite{M} for a more conceptual proof in the context of Lie algebras.

\begin{theorem}\label{spitzer}
Let $(\A,R,\A_n)$ be a complete filtered Rota-Baxter algebra of
weight $\theta=1$. Let $a \in \A_1$.
\begin{enumerate}
\item
The equation
\begin{equation}
 b=1-R(ba)
\mlabel{eq:recurs}
\end{equation}
has a unique solution
\begin{equation}
b= \exp\big(-R(\chi(\log (1+a)))\big). \mlabel{eq:exp}
\end{equation}
\item
The equation
\begin{equation}
b=1-\tilde{R}(ab) \mlabel{eq:recurs2}
\end{equation}
has a unique solution
\begin{equation}
b= \exp\big(-\tilde{R}(\chi(\log (1+a)))\big). \mlabel{eq:exp2}
\end{equation}
\end{enumerate}
\mlabel{thm:equation}
\end{theorem}

\begin{proof}
We only need to verify for the first equation. The proof for the
second equation is similar.

Since $a$ is in $A_1$ and $R$ preserves the filtration, the series
$$
b=1+ R(a)+ R(R(a)a) + \cdots +(Ra)^{[n]} + \cdots
$$
defines a unique element in $\A$ and is easily seen to be a
solution of (\ref{eq:recurs}). Conversely, if $c \in \A$ is a
solution of (\ref{eq:recurs}), then by iterated substitution, we
have
$$
c=1+R(a)+R(R(a)a)+ \cdots +(Ra)^{[n]}+\cdots.
$$
Therefore, the equation (\ref{eq:recurs}) has a unique solution.

To verify that (\ref{eq:exp}) gives the solution, take $u:=\log
(1+a), \; a\in \A_1$. Using (\ref{eq:Kingman}), for our chosen $b$
we have \allowdisplaybreaks{
\begin{eqnarray*}
\lefteqn{\exp\big(-R(\chi(\log (1+a)))\big)= \exp\big(-R(\chi(u))\big) }\\
&=& \sum_{n=0}^\infty \frac{\big(-R(\chi(u))\big)^n}{n!} \\
&=& 1 + R \bigg(\sum_{n=1}^\infty \frac{(-1)^n}{n!}\bigg
(\big(R(\chi(u))\big)^n
            -\big(-\tilde{R}(\chi(u))\big)^n\bigg)\bigg)\\
&=& 1 + R \bigg(\sum_{n=0}^\infty
\frac{(-1)^n}{n!}\big(R(\chi(u))\big)^n
            -\sum_{n=0}^\infty \frac{(-1)^n}{n!}\big(-\tilde{R}(\chi(u))\big)^n\bigg)\\
&=& 1 +
R\bigg(\exp\big(-R(\chi(u))\big)-\exp\big(\tilde{R}(\chi(u))\big)\bigg).
\end{eqnarray*}}
By the definition of the $BCH$-recursion $\chi$ in equation
(\ref{BCH-recur}), we have
\allowdisplaybreaks{\begin{eqnarray*}
&&\exp\big(R(\chi(u))\big)\exp\big(\tilde{R}(\chi(u))\big) \\
&=&\exp\bigg(R\big(\chi(u)\big)+\tilde{R}\big(\chi(u)\big)
            +BCH\big(R(\chi(u)),\tilde{R}(\chi(u))\big)\bigg)\\
&=& \exp \bigg( \chi(u)+BCH\big(R(\chi(u)),\tilde{R}(\chi(u))\big)\bigg)\\
&=& \exp(u).
\end{eqnarray*}}
Thus
\allowdisplaybreaks{\begin{eqnarray*}
&&\exp\big(-R(\chi(\log (1+a)))\big)\\
&=& 1+R\bigg (\exp\big(-R(\chi(u))\big)-\exp\big(-R(\chi(u))\big)\exp(u)\bigg)\\
&=& 1+R\bigg (\exp\big(-R(\chi(u))\big)\big(1-\exp(u)\big)\bigg)\\
&=& 1+R\bigg (\exp\big(-R(\chi(\log(1+a)))\big)\big(1-\exp(\log(1+a))\big)\bigg )\\
&=& 1-R\bigg (\exp\big(-R(\chi(\log(1+a))\big)\;a\bigg)
\end{eqnarray*}}
This verifies the first equation.
\end{proof}

\begin{corollary}
Let $(\A,R,\A_n)$ be a complete filtered Rota-Baxter algebra of
weight $\theta=1$. For $a \in \A_1$, we have
\begin{equation}
\sum_{n=0}^\infty \big(Ra\big)^{[n]}
    = \exp\big(-R(\chi(\log (1+a)))\big)
\mlabel{eq:Spitzer1}
\end{equation}
\begin{equation}
\sum_{n=0}^\infty \big(\tilde{R}a\big)^{\{n\}}
    = \exp\big(-\tilde{R}(\chi(\log (1+a)))\big)
\mlabel{eq:Spitzer2}
\end{equation}
\mlabel{co:Spitzer}
\end{corollary}

\begin{proof}
By Theorem~\ref{thm:equation} and its proof, both sides of
(\ref{eq:Spitzer1}) are solutions of (\ref{eq:recurs}). This
proves (\ref{eq:Spitzer1}).

The proof of (\ref{eq:Spitzer2}) is the same, by considering
solutions of the recursive equation (\ref{eq:recurs2})
\end{proof}

For later reference, we record here a simple and attractive
theorem of Atkinson~\cite{A} whose proof just uses relations
(\ref{ab}) and (\ref{ba}).

\begin{theorem}
Let $(\A,R)$ be an associative unital but not necessarily
commutative Rota-Baxter algebra. Assume $b$ and $b'$ to be
solutions of the recursive equations (\ref{eq:recurs}) and
(\ref{eq:recurs2}), then
$$ b (1+a) b' = 1.$$
\label{Atkinson2}
\end{theorem}

We now prove the Birkhoff decomposition of filtered Rota-Baxter
algebras.

\begin{theorem}
Let $(\A,R)$ be an associative unital complete Rota-Baxter algebra
with filtration $A_n,\ n\geq 0$. The following conditions are
equivalent.
\begin{enumerate}
\item[(i)] $R$ is idempotent: $R^2=R$ when restricted to $A_1$.
\item[(ii)] There is a direct product decomposition of algebras
$$ A_1 = R(A_1) \times \tilde{R}(A_1).$$
\item[(iii)] There is a direct product decomposition of groups
$$ (1+A_1)= (1+R(A_1)) \times (1+\tilde{R}(A_1)).$$
\end{enumerate}
\mlabel{thm:Birkhoff}
\end{theorem}
\begin{rmk}
Under the assumption in ($i$), the statement in ($ii$) is the
Atkinson decomposition \cite{A} and the statement in ($iii$)
specializes to give the uniqueness of the Birkhoff decomposition
of Connes and Kreimer. See Section \ref{Birkhoff} for details.
\end{rmk}

\begin{proof}
(1) $\Leftrightarrow$ (2) is clear and does not need the
completeness assumption.

(2) $\Rightarrow$ (3): We just need to show that, for each $a\in
A_1$, there is a unique $c\in R(A_1)$ and a unique $\tilde{c}\in
\tilde{R}(A_1)$ such that
$$ 1+a = (1+c)(1+\tilde{c}).$$

Let $a\in A_1$ be given, and let $b$ and $\tilde{b}$ be the
solution of (\ref{eq:recurs}) and (\ref{eq:recurs2}) respectively.
Then by Theorem~\ref{Atkinson2}, we have
$$ b(1+a)\tilde{b}=1.$$
By their constructions and (\ref{Rhom1},\ref{Rhom2}), we have
$b=1-b_1$ and $\tilde{b}=1-\tilde{b}_1$ for $b_1\in R(A_1)$ and
$\tilde{b}_1\in \tilde{R}(A_1)$. Thus
$$ b^{-1} =1+b_1+b_1^2+\cdots \in 1+ R(A_1),$$
$$ \tilde{b}^{-1} = 1+\tilde{b}_1+\tilde{b}_1^2+ \cdots \in 1+\tilde{R}(A_1).$$
This proves the existence.

For the uniqueness, suppose we have
$$ 1+a = (1+c)(1+\tilde{c}) = (1+d) (1+\tilde{d})$$
with $c,d\in R(A_1)$ and $\tilde{c},\tilde{d}\in \tilde{R}(A_1)$.
Then
$$ (1+d)^{-1}(1+c) = (1+\tilde{d})(1+\tilde{c})^{-1}$$
which is in $(1+R(A_1)) \cap (1+\tilde{R}(A_1))$. But this
intersection is $\{1\}$ because
$$1+R(d)=1+\tilde{R}(d') \Rightarrow R(d)=\tilde{R}(d')\Rightarrow R(d)=0.$$

(3) $\Rightarrow$ (2): Since $R+\tilde{R}=\id$, we have $A_1 =
R(A_1)+\tilde{R}(A_1)$. So we just need to show $R(A_1)\cap
\tilde{R}(A_1)=0$. This is true if and only if $(1+R(A_1))\cap
(1+\tilde{R}(A_1))=\{1\}.$.
\end{proof}

\subsection{Algebraic Bogoliubov map}

For $a \in A_1$, let $a_-$ be the unique solution of $b=1-R(ba)$
from Theorem~\ref{thm:equation} and let
$$
\gamma(a)=a_-\, a.
$$
Similarly, let $\tilde{a}$ be the unique solution of $
b=1-\tilde{R}(ab)$ and let
$$
\tilde{\gamma}(a)=a\tilde{a}.
$$
By Proposition~\ref{prop-double}, $A_1$ with the
product $\ast_R$ is still a complete algebra.
Define
$$
\exp_R: A_1 \to 1+A_1,\ \exp_R(a) := \sum_{n=0}^\infty
\frac{a^{\ast_R n}}{n!}
$$
where $a^{\ast_R n}$ is the $n$-th power of $a$ under the product
$\ast_R$.

\begin{theorem}
The following diagram commutes.
\begin{equation}
\xymatrix{ A_1 \ar[rr]^{\exp} \ar[dd]_{-\chi} & & 1+A_1
\ar[dd]^{\beta}
    \ar[rr]^{\theta} && A_1 \ar[dd]^{-\gamma}\\
&&\\
A_1  \ar[rr]^{\exp_R} \ar[dd]_{R \times (-\tilde{R})}
    & & 1+A_1 \ar[dd]^{R'\times (-\tilde{R}')}
    \ar[rr]^{\theta} && A_1 \ar[dd]^{R\times (-\tilde{R})} \\
&& \\
R(A_1) \times \tilde{R}(A_1) \ar[rr]^{\exp\times \exp}
            & & (1+R(A_1)) \times (1+\tilde{R}(A_1))
    \ar[rr]^{\theta\times \theta} && R(A_1) \times \tilde{R} (A_1)
} \mlabel{eq:diag}
\end{equation}
Here $\theta(x)=x-1$ and $\beta$ is defined to be the composite
$$\beta = \theta^{-1} \circ (-\gamma) \circ \theta. $$
So $\beta(c)=1-\gamma(c-1)$. Similarly define $R'\times
(-\tilde{R}')$. So
 \allowdisplaybreaks{\begin{eqnarray*}
 R': && 1+A_1 \to 1+A_1,\ a\mapsto 1+R(a-1),\\
\tilde{R}': && 1+A_1 \to 1+A_1,\ a\mapsto 1-\tilde{R}(a-1).
\end{eqnarray*}}
\mlabel{thm:algCK}
\end{theorem}
We call the map $\beta$ the algebraic Bogoliubov because it gives the
Bogoliubov map in renormalization theory.

\begin{proof}
We only need to prove the commutativity of the upper half and the
lower half of the diagram. By the way the two maps in the middle
column are defined and by the bijectivity of the horizontal maps
in the right half of the diagram, it follows that the other squares are
also commutative.

Verifying the commutativity of the top half means to verify
$$
-\gamma \circ \theta \circ \exp (u)=\theta \circ \exp_R \circ
(-\chi(u)),
$$
that is,
$$
-\gamma (\exp(u)-1) = \exp_R(-\chi(u))-1.
$$
By Theorem~\ref{thm:equation}, and reversing the derivations in
its proof, we have \allowdisplaybreaks{\begin{eqnarray*}
 -\gamma (\exp(u)-1) &=& -\exp\big(-R(\chi(u)\big)\big(\exp(u)-1\big)\\
&=& -\exp\big(-R(\chi(u))\big)\exp(u)-\exp\big(-R(\chi(u))\big)\\
&=& \exp\big(-R(\chi(u))\big) -\exp\big(\tilde{R}(\chi(u))\big).
\end{eqnarray*}}
By (\ref{Rhom1}) and (\ref{Rhom2}) we obtain
 \allowdisplaybreaks{\begin{eqnarray}
R\big(\exp_R(u)\big)&=&\exp\big(R(u)\big)+R(1) -1, \mlabel{eq:double1}\\
\tilde{R}\big(\exp_R(u)\big) &=&
-\exp\big(-\tilde{R}(u)\big)+\tilde{R}(1)+1. \mlabel{eq:double2}
\end{eqnarray}}
Thus the last term of the earlier equation is
 \allowdisplaybreaks{\begin{eqnarray*}
&& R\big(\exp_R(-\chi(u))\big)+1-R(1)+\tilde{R}\big(\exp_R(-\chi(u))\big)-1-\tilde{R}(1)\\
&=& \exp_R\big(-\chi(u)\big) - 1,
\end{eqnarray*}}
as is desired.

Verifying the commutativity of the lower half of the diagram means
to verify the two equations
$$
\exp\big(R(u)\big)-1=R\big(\exp_R(u)-1\big),\quad
\exp\big(-\tilde{R}(u)\big)-1
    = -\tilde{R}\big(\exp_R(u)-1\big)
$$
which are immediately from (\ref{eq:double1}) and
(\ref{eq:double2}).
\end{proof}

\subsection{General Rota-Baxter algebras}

Now let $(\A,R)$ be any Rota-Baxter algebra of weight $\theta=1$.
Consider the power series ring $\A[[x]]$ on one (commuting)
variable x. So $\A[[x]]=\Z[[x]]\otimes \A$. Define an operator,
$$
\RB: \A[[x]] \to \A[[x]],\ \RB(\sum_{n=0}^\infty a_n x^n)=\sum_{n=0}^\infty R(a_n)x^n.
$$

\begin{lemma}
$(\A[[x]],\RB)$ is a Rota-Baxter algebra.
\end{lemma}
\begin{proof}
This is a straight forward verification. For $f=\sum_n a_n x^n$,
$g=\sum_m b_m x^m$, we have
\allowdisplaybreaks{
\begin{eqnarray*}
\RB(f)\RB(g)&=& \big(\sum_n R(a_n) x^n\big ) \big(\sum_m R(b_m) x^m \big)\\
&=& \sum_{n,m} R(a_n)R(b_m) x^{m+n}\\
&=& \sum_{n,m} \big( R(R(a_n)b_m)+R(a_nR(b_m))-R(a_nb_m)\big) x^{m+n}\\
&=& \RB ((\sum_n R(a_n)x^n)(\sum_m b_mx^m))
    +\RB ((\sum_n a_n x^n)(\sum_m R(b_m) x^m))\\
&&    -\RB((\sum_n a_n x^n)(\sum_m b_m x^m))\\
&=& \RB(\RB(f) g)+\RB(f\RB(g))-\RB(fg).
\end{eqnarray*}}
\end{proof}

Now it is easy to verify that, with the filtration
$$
\A_n:= x^n \A[[x]],\ n \geq 0,
$$
$\A[[x]]$ is a complete Rota-Baxter algebra. By
Theorem~\ref{spitzer}, we have
\begin{corollary}
For $a \in \A$, we therefore have
\begin{equation}
\sum_{n=0}^\infty (\RB(ax))^{[n]}
    = \exp\big(-\RB(\chi(\log (1+ax)))\big)
\mlabel{eq:Spitzer3}
\end{equation}
\end{corollary}

\begin{rmk}
{\rm{ Obviously, for $\A$ being commutative, we have $\chi(a)=a$,
and relation (\ref{eq:Spitzer1}) just reduces to the classical
Spitzer's identity. Our result therefore is the natural
non-commutative generalization of this well-known identity.}}
\end{rmk}

By comparing coefficients of similar powers of $x$ on the two
sides of the equation (\ref{eq:Spitzer3}), we obtain identities in
Rota-Baxter algebras that are not necessarily commutative.

\section{Birkhoff decomposition in renormalization theory} \label{Birkhoff}

Now we consider the case when the complete Rota-Baxter algebra is
as in Example~\ref{FG}. We will use the notations in articles
\cite{EGK1} and \cite{EGK2}. For a general review on the Hopf
algebraic approach to renormalization theory in pQFT, we refer the
reader to the original work \cite{CK3, CK1, Kreim1, Kreim2}. For a
recent and elaborate review of the Connes-Kreimer work on
renormalization theory, we refer the reader to the work by Manchon
\cite{M}.

Kreimer and later Connes and Kreimer were able to uncover the
mathematical content underlying the algebraic combinatorial
process of renormalization theory in pQFT, by organizing the
combinatorics in terms of a combinatorial, i.e.\ graded connected
Hopf algebra structure on Feynman graphs, denoted by $\H_{FG}$.
Furthermore, by interpreting Feynman rules as regularized
characters, i.e.\ multiplicative maps from the above Hopf algebra
of Feynman graphs into an associative unital and commutative
Rota-Baxter algebra, the process of renormalization became a
Birkhoff decomposition of these characters.

We will denote the space of linear functionals from $\H_{FG}$ into
the Rota-Baxter algebra $(\A,R)$ by $L(\H_{FG},\A)$.
$L(\H_{FG},\A)$ carries the structure of an associative unital
non-commutative algebra with respect the convolution product,
denoted by
$$
f \star g := m_{\A}(f \otimes g)\Delta, \; f,g\in L(\H_{FG},\A).
$$
Here $\Delta$ denotes the coproduct in $\H_{FG}$. The unit in
$L(\H_{FG},\A)$ is given by the counit $\epsilon:\H_{FG} \to 1\K$.
Let $\phi$ be a regularized character, i.e. an element in the
group $G \subset L(\H_{FG},\A)$, generated by the infinitesimal
characters forming a Lie algebra $g \subset L(\H_{FG},\A)$. We
then lift the Rota-Baxter map $R: \A \to \A$ to the algebra
$L(\H_{FG},\A)$, see Proposition (\ref{lift}) below.

In \cite{CK1}, it was shown that for arbitrary $\phi \in G$ there
exist two unique characters, defined recursively for $\Gamma \in
ker(\epsilon)\subset \H_{FG}$ by
\allowdisplaybreaks{
\begin{eqnarray}
\phi_{\pm}: \H_{FG} \to \A,
\begin{cases}
\phi_{-}(\Gamma) := -R\big[\phi(\Gamma) +
\sum_{(\Gamma)}'\phi_{-}(\Gamma')\phi(\Gamma'')\big], \label{SR} \\
\phi_{+}(\Gamma) := \tilde{R}\big[\phi(\Gamma) +
\sum_{(\Gamma)}'\phi_{-}(\Gamma')\phi(\Gamma'')\big]
\label{renorm}, & \makebox{ and}\\
\phi_{\pm}(1):=1
\end{cases}
\end{eqnarray}}
such that
\begin{equation}
\phi=\phi^{-1}_{-} \star \phi_{+}. \label{CK}
\end{equation}
Here we used Sweedler's notation, $\Delta(\Gamma):=\Gamma\otimes 1
+ 1 \otimes \Gamma + \sum_{(\Gamma)}' \Gamma' \otimes \Gamma''$
for $\Gamma \in \H_{FG}$. The character $S_R^{\phi}:=\phi_{-}$ was
called twisted antipode, and provides the counterterm. The
so-called renormalized character $\phi_{+}$ gives the renormalized
Feynman rules. To proof the multiplicativity of $\phi_{-}$ and
$\phi_{+}$ essential use of the Rota-Baxter structure on the
target space $\A$ of the characters was made. Using the following

\begin{proposition} {\rm \cite{EGK1}} \label{lift}
Define the linear map $\RB: L(\H_{FG},\A) \to L(\H_{FG},\A)$ by $f
\mapsto \RB(f):=R \circ f: \H_{FG} \to R(\A)$. Then
$L(\H_{FG},\A)$ becomes an associative, unital non-commutative
Rota-Baxter algebra. The Lie algebra of infinitesimal characters
$g \subset L(\H_{FG},\A)$ becomes a Lie Rota-Baxter algebra, i.e.
for $Z',Z'' \in g$,
\begin{equation}
    [\RB(Z'),\RB(Z'')]=\RB\big([Z',\RB(Z'')]\big) + \RB\big([\RB(Z'),Z'']\big) - \RB \big([Z',Z'']\big).
    \label{Lie}
\end{equation}
\end{proposition}
We can write equivalently, $\phi_{-}$ in terms of the recursive
equation
\begin{equation}
\phi_{-} = \epsilon - \RB\big[\phi_{-} \star (\phi \circ J)\big],
\label{phi-}
\end{equation}
where $J$, the projector onto the augmentation ideal
$ker(\epsilon)$, is defined in terms of the unit map $\eta:1\K \to
\H_{FG}$, $J:=id_{\H_{FG}} - \eta\epsilon$. Note that by linearity
of $\phi$ we have
$$ (\epsilon + \phi \circ J)= \phi.$$
Let $\phi \in G$ be generated by $Z \in g$, i.e.
$\phi=\exp^{\star}(Z)$. So by Theorem~\ref{spitzer}, the recursion
(\ref{phi-}) for $\phi_{-}$ is solved by \allowdisplaybreaks{
\begin{eqnarray}
\phi_{-} &=& \exp^\star\big(-\RB\big(\chi(log^{\star}(\epsilon +
\phi\circ J))\big)\big) \\
         &=& \exp^\star\big(-\RB\big(\chi(Z)\big)\big)
\end{eqnarray}}
as proved in \cite{EGK2}.

We now let $\tilde{\phi}$ be defined by the recursive equation
\allowdisplaybreaks{
\begin{equation}
\tilde{\phi} = \epsilon - \tilde{\RB}\big[(\phi \circ J) \star
\tilde{\phi}\:\big]. \label{eq:phi+}
\end{equation}}
So by Theorem~\ref{Atkinson2}, we have
$$
\phi_{-} \star \phi \star \tilde{\phi} = \epsilon.
$$
On the other hand, following (\ref{CK}), it is well-known that,
for the unique renormalized character $\phi_+$, we have
$$
\phi_{-} \star \phi \star \phi_{+}^{-1} = \epsilon.
$$
Since both equations hold in the Lie group $G$ of regularized
characters, we must have
\allowdisplaybreaks{
\begin{eqnarray}
\tilde{\phi} &=& \phi_{+}^{-1}  \\
              &=&  \exp^{\star}\big( -\tilde{\RB}(\chi(Z))\big).
              \label{expo}
\end{eqnarray}}
The second equality follows by Theorem~\ref{spitzer} equation
(\ref{eq:exp2}) and was shown for $\phi_{+}$ directly in
\cite{EGK2}. This simple result implies a new recursive relation
for $\phi_{+}$ in terms of $\tilde{\RB}$ \allowdisplaybreaks{
\begin{equation}
\phi_{+} = \epsilon - \tilde{\RB}\big[\phi_{+} \star (\phi^{-1}
\circ J)\big].
\end{equation}}

Note that this result is completely natural. The antipode $S$
($S^2={\rm id}$) can be written in terms of the projector $J$ as
\begin{equation}
S=-m \circ (S \otimes J)\circ\Delta=-m\circ (J \otimes
S)\circ\Delta.
\end{equation}
Iterating $\phi_-$ on the left hand side of the tensor product, it was used
to deform the character $\phi\circ S$ to the counterterm character
$\phi_-$. But one naturally expects that one also can derive the
forest formula by recursing $\phi_+$, and this is what the above
formula achieves. The appearance of $\phi^{-1}$ then instead of
$\phi$ compensates for the minus sign in front of $\tilde{\RB}$,
making use of the very exponentiation in (\ref{expo}).

Also, we remind ourselves that the renormalized character is a
character in the image of $\tilde{\RB}$, where, $\tilde{\RB}$ acts
on the Bogoliubov character, a map which replaces all
subdivergences by their evaluation under $\phi_+$, a fact
guaranteed by the structure of the Hochschild cohomology of such
Hopf algebras \cite{Bergbau, Houches}. Thus, one naturally
recurses $\phi_+$ in terms of itself, a fact evident also in the
basic structure of renormalized Dyson--Schwinger equations, which
can be completely written in terms of themselves. The above
formula makes that fact self-evident on a combinatorial level.

Summarizing, by the above argument we find naturally the following
two recursions for the factors of the Birkhoff decomposition of a
character $\phi$:
\allowdisplaybreaks{
\begin{equation}
\phi_{+} = \epsilon -\tilde{\RB}\big[\phi_{+} \star (\phi^{-1}
\circ J)\big] \;\;\makebox{ and }\;\; \phi_{-} = \epsilon
-\RB\big[\phi_{-} \star (\phi \circ J)\big].
\end{equation}}
Using the augmentation ideal projector \makebox{$J:=id_{\H_{FG}} -
\eta\epsilon$} we can derive the simple identity
\allowdisplaybreaks{
\begin{eqnarray}
\phi \star (\phi^{-1} \circ J) &=& \phi \star \phi \circ S \circ J \\
                         &=& \phi \star \phi \circ S \circ (id_{\H_{FG}} -
                                                            \eta\epsilon)\\
                         &=& \phi \star \phi \circ S - \phi \star (\phi \circ
                                                        S \circ \eta\epsilon)\\
                         &=& \epsilon - \phi = - \phi \circ J,
\end{eqnarray}}
which allows us to show, using
\makebox{$\phi=\phi_{-}^{-1}\star\phi_{+}$}, that
\allowdisplaybreaks{
\begin{eqnarray}
-\phi_{+} \star (\phi^{-1} \circ J) &=& \phi_{+} \star \phi^{-1} \star (\phi \circ J) \\
                         &=& \phi_{+} \star \phi_{+}^{-1} \star \phi_{-} \star (\phi \circ J) \\
                         &=& \phi_{-} \star \phi \circ J. \mlabel{eq:+-}
\end{eqnarray}}
This allows us to get back the original $\phi_{+}$-recursion
(\ref{renorm}) in terms of the Bogoliubov character \cite{EGK2},
i.e. Bogoliubov's R-map, defined via the double product
$\star_\RB$, $\exp^{\star_{\RB}}(\chi(Z))=\phi_{-} \star \phi
\circ J$
\begin{equation}
\phi_{+} = \epsilon + \tilde{\RB}\big[\phi_{-} \star (\phi \circ
J)\big].
\end{equation}

Let us for the sake of clarity compare the above results in the
setting of combinatorial Hopf algebras and regularized characters
with the findings of Section 3, i.e. general filtered Rota-Baxter
algebras. To clearly show the connection, we display the following
''dictionary". We fix a character $\phi: \H_{FG} \to \A$ in $G$
and let $a:=\phi \circ J$. In the following table, entries in the
left column are results proved earlier in this paper for general
complete filtered Rota-Baxter algebras, and entries in the right
column are their interpretations in the non-commutative
associative unital Rota-Baxter algebra $(L(\H_{FG},\A), \RB)$.

$$
 \allowdisplaybreaks{
\begin{array}{lll}
a                  & & \phi\circ J \\[0.1cm]
a_-=1 - R(a_-\, a) & & \phi_- =  \ep - \RB\big(\phi_- \star (\phi\circ J)\big)\\[0.1cm]
\tilde{a}=1 - \tilde{R}(a \tilde{a}) & &
\tilde{\phi} = \ep - \tilde{\RB}\big((\phi\circ J) \star \tilde{\phi}\big)\\[0.1cm]
a_-\, (1+a) \tilde{a} = 1 & & \phi_- \star (\ep + \phi\circ J)
\star \tilde{\phi}
        =\phi_- \star \phi \star \tilde{\phi} =\ep\\[0.1cm]
a_+:=\tilde{a}^{-1} = a_-\,(1+a) && \phi_+:= \phi_- \star \phi = \tilde{\phi}^{-1}\\[0.1cm]
a_+ \stackrel{(i)}{=} 1-\tilde{R}\big(a_+ (\frac{-a}{1+a})\big)
&& \phi_+=\ep -\tilde{\RB}\big(\phi_+ \star (\phi^{-1}\circ J)\big)\\[0.1cm]
-a_+ \big(\frac{-a}{1+a}\big)\stackrel{(ii)}{=}a_-\, a && -\phi_+
\star (\phi^{-1}\circ J) =\phi_- \star (\phi\circ J)
\end{array}}
$$

\begin{proof}
(i) \allowdisplaybreaks{
\begin{eqnarray}
1-\tilde{R}\big(a_+ (\frac{-a}{1+a})\big) &=&
1-\tilde{R}\big(a_-\, (1+a) (\frac{-a}{1+a})\big) \\
&=& 1+\tilde{R}(a_-\, a)=1-R(a_-\, a)+a_-\, a= a_+
\end{eqnarray}}

(ii) \allowdisplaybreaks{
\begin{eqnarray}
-a_+ \frac{-a}{1+a} &=& - a_-\ (1+a) \frac{-a}{1+a} \\
                    &=& a_{-}a
\end{eqnarray}}
\end{proof}

The diagram (\ref{eq:diag}) specializes to the following diagram
in the case of renormalization: Let $g$ be the complete filtered
Lie algebra of derivations in $L(\H_{FG},\A)$, $G$ the Lie group
of characters. For $\Gamma \in \H_{FG}$, let $b[\phi](\Gamma) =
-(\phi(\Gamma)+\sum'_{(\Gamma)} \phi_-(\Gamma')\phi(\Gamma''))$ be
the Bogoliubov map. We have the following diagram when restricted
to $\ker \ep$
\[\xymatrix{
g \ar[rr]^{\exp^\star} \ar[dd]_{-\chi} & & G \ar[dd]^{b} \\
&&\\
g \ar[rr]^{\exp^{\star_\RB}} \ar[dd]_{\RB \times (\RB-\id)}
    & & G_\RB \ar[dd]^{\RB \times (\RB-\id)} \\
&& \\
g^- \times g^+ \ar[rr]^{\exp^\star} & & G^- \times G^+ }\] 
This is the reason that the map $\beta$ in (\ref{eq:diag}) is called
the algebraic Bogoliubov map.

\section{Summary and Outlook}

In this work we derived, in the realm of complete filtered
Rota-Baxter algebras, by simple algebraic terms a non-commutative
version of Spitzer's identity. The latter is a well-known object
in the theory of random variables. The simplicity of the proofs
relies on a more general result obtained in previous work by
solving the recursively defined formulae of the Birkhoff
decomposition of regularized characters  in terms of a co-called
$BCH$-recursion. Initially, this was done in the Connes-Kreimer
Hopf algebraic approach to renormalization theory in pQFT. This
approach allowed us to derive a new forest-like formula for the
renormalized character.

Also, we believe that the fact that the classical Spitzer's
formula is intimately related to theory of symmetric functions and
generalizations of the shuffle product might allow us to extend
these connections via its non-commutative version given here.\\[0.3cm]

{\emph{Acknowledgements}}: The first author warmly thanks the Ev.
Studienwerk for financial support, and the Institut Henri
Poincar\'e for hospitality.



%

\end{document}